\documentstyle[aps,epsf,preprint]{revtex}
\def\be{\begin{equation}}
\def\ee{\end{equation}}
\def\bearr{\begin{eqnarray}}
\def\eearr{\end{eqnarray}}

\begin{document}
\draft
\preprint{}

\title{Meson Correlators at Finite Temperature}
\author{
Varun Sheel\footnote {Electronic address: varun@prl.ernet.in}, 
Hiranmaya Mishra 
and Jitendra C. Parikh } 
\address{Theory Division, 
Physical Research Laboratory, Navrangpura, Ahmedabad 380 009, India}
\maketitle
\begin{abstract}
We evaluate equal time point to point spatial correlation 
functions of mesonic currents at finite temperature.
For this purpose we consider the QCD vacuum structure in terms of quark
antiquark condensates and  their fluctuations in terms of an
irreducible four point structure of the vacuum.
The temperature dependence of quark
condensates is modeled using chiral perturbation theory for 
low temperatures and lattice QCD simulations near
the critical temperature.We first consider the propagation of
quarks in a condensate medium at finite temperature.
We then determine the correlation functions
in a hot medium. Parameters such as mass, coupling constant and
threshold energy are deduced from the finite temperature
correlators. We find that all of them decrease close to the
critical temperature.
\end{abstract}
\vskip 0.5cm
 \section{Introduction}
 The structure of vacuum in
Quantum Chromodynamics (QCD) is one of the most interesting
question in strong interaction physics \cite{sur}. The evidence for
quark and gluon condensates in vacuum is a reflection of its complex nature
\cite{svz}. Determination of correlation functions \cite{surcor,neglecor}
of hadronic currents in such a vacuum state provides rich 
information regarding interquark interaction as a
function of their spatial separation as well as  on
hadron spectroscopy. These are some of the nonperturbative
feature of QCD and are of great value in understanding
the ground state structure of the theory of strong 
interactions\cite{surcor,neglecor}.

We have studied  mesonic and baryonic current correlators at zero temperature 
with a non-trivial structure for the ground state with quark antiquark
condensates\cite{prliop,propcor}. It was shown that the square of the 
quark propagator does not reproduce the correlation function for the pion
deduced from phenomenology. In order to match the data it was necessary to 
introduce an irreducible four point structure for the quarks in the vacuum.
This may be looked upon as an effective way of incorporating 
gluon condensate contribution to the correlator.

As is well known \cite{lman} the QCD vacuum state changes with temperature. 
Lattice monte carlo simulations suggest that chiral symmetry is restored
around 150 MeV. In view of this the present note is
aimed at looking at the behaviour of the meson correlation
functions at finite temperature. This is of great interest in the
context of behaviour of hadrons around the chiral phase transition
associated with quark gluon plasma\cite{hotcor,shuhotcor}. It may be 
noted that there is
little phenomenological information in this regime 
but there are several theoretical studies \cite{sumt} essentially using
sum rule methods. The main objective here is to employ a different
nonperturbative approach developed by us \cite{plb}. This has been sucessful 
at zero temperature, and, its extension to finite temperature  is 
therefore of interest. In particular, we will obtain temperature dependence 
of masses, coupling constants and threshold energies for the pion 
and rho mesons.

We organise the paper as follows. 
In section II we discuss the quark condensate at finite
temperature to fix the parameter appearing in the ansatz of the
ground state of QCD.  We then discuss in section III the
quark propagation in the thermal vacuum.
In section IV we calculate meson correlation functions
at finite temperature. Finally we  discuss the results in section V.

\section{Quark Condensate at Finite Temperature}
To calculate the correlators at finite temperature 
we need the expression for the equal time propagator for the
interacting quark field operators. 
We have developed earlier\cite{prliop} a vacuum structure in terms
of quark antiquark condensates with a condensate function $h(k)$.
The equal time propagator could then be calculated in terms of the
condensate function\cite{propcor}. One can generalise this to
finite temperature using the method of thermofield dynamics. Here
the thermal average is obtained as an expectation value of
the operator over the thermal vacuum\cite{umezawa}. This leads to
\bearr
\langle \psi^i_\alpha(\vec x)\psi^{j\dagger}_\beta(\vec
0)\rangle_{T}&=&\frac{\delta^{ij}}{(2\pi)^3}\int e^{i\vec
k\cdot\vec x} \Lambda_{+\alpha\beta}(\vec k,T)d\vec k\nonumber\\
\langle \psi^{i\dagger}_\alpha(\vec x)\psi^{j}_\beta(\vec
0)\rangle_{T}&=&\frac{\delta^{ij}}{(2\pi)^3}\int e^{-i\vec
k\cdot\vec x} \Lambda_{-\beta\alpha}(\vec k,T)d\vec k
\label{psi}
\eearr
 The thermal vacuum is obtained
from the zero temperature vacuum by a thermal Bogoliubov
transformation in an extended Hilbert space involving extra field
operators (thermal doubling of operators)\cite{umezawa}.
The functions $\Lambda_{\pm}$ (Eq.~\ref{psi}) 
for the case of two flavour
massless quarks, are given as(with $k=|\vec k|$),
\be
\Lambda_{\pm}(\vec k, T)=\frac{1}{2}\bigg[ 
1\pm \cos 2\theta(\gamma^0 \sin\!2h(k) + \vec\alpha\cdot\hat k 
\cos\!2h(k))\bigg].
\label{lpm}
\ee

\noindent In the above h(k) is the
 condensate function\cite{prliop,propcor,plb}
corresponding to the Bogoliubov transformation to include a condensate
structure in the vacuum. The function $\theta$ is  associated with 
the thermal Bogoliubov transformation and
is related to the distribution function as\cite{umezawa}
\begin{equation}
\sin^2\theta (k)=\frac{1}{\exp[\beta\epsilon (k)]+1}\!,
\ee
$\beta$  being the inverse temperature. 
Further $\epsilon(k)$ is the single particle energy
given as $\epsilon(k)=\sqrt{k^2+m(k)^2}$. In the presence of 
condensate  the dynamical mass is given as
 $m(k)=m+k \tan 2h(k)$, $m$ being a possible 
current quark mass\cite{prliop}.

We had earlier taken a gaussian ansatz for the condensate function
$\sin\! 2h(k)=e^{-R^2 k^2/2}$.
In order to determine the parameter $R$, we had taken a 
value of $R$ consistent with hadronic  correlator phenomenology. We
 choose a similar structure for the condensate function
at finite temperature namely $ \sin\! 2h(k)=e^{-R(T)^2k^2/2} $
with $R(T)$ now being temperature dependent. 

In order to determine $R(T)$ or equivalently the ratio
$S(T)=R(T=0)/R(T)$, we
first evaluate our expression of the order parameter 
(the condensate value) at finite temperature. In terms of the
dimensionless variable $\eta=Rk$, this is given as
\be
\frac{\langle \bar q q\rangle_T}{\langle \bar q q\rangle_{T=0}}
= S(T)^3\bigg[ 1-2\sqrt{\frac{2}{\pi}}\int
e^{-\eta^2/2} \sin^2(z,\eta) \eta^2 d\eta \bigg],
\label{qqrat}
\ee
where $\sin^2\theta(z,\eta)=\frac{1}{e^{z\epsilon(\eta)}+1}$, with 
$z=\beta/R(T)$ and $\epsilon(\eta)=\eta/\cos 2h(\eta)$.

We can obtain  $S(T)=R(T=0)/R(T)$ if we know the temperature
dependence of the order parameter
on the left hand side of Eq.(\ref{qqrat}).
As there are no phenomenological inputs for this, we shall consider the
results from chiral perturbation theory (CHPT) which is expected 
to be valid at least for
small temperatures. For higher temperatures near the critical 
temperature, lattice simulations seem to yield the universal 
behaviour\cite{lman} with a large correlation length associated with 
a second order phase transition
for two flavor massless QCD. We shall use such a critical behaviour to
consider the temperature dependence of the order parameter near the 
critical temperature.

We quote here the results of CHPT
obtained by Gerber and Leutwyler\cite{chpt}.
The condensate ratio at temperatures small compared to the pion
mass is given as
\be
\frac{\langle \bar q q\rangle_T}{\langle \bar q q\rangle_{T=0}}
=1+\frac{c}{F^2} \left [ 
\frac{3}{2} T^4 h_0^{'} + 4\pi T^4 \left(a' h_1^2 + 2 a h_1 h_1^{'}
\right) + \pi T^8 \left(h' b_{eff} + b_{eff}^{'} h \right)
\right ].
\label{qqchp}
\ee
where the functions $h$ are defined as 
\bearr
h_0=H^4(\mu)/(3\pi^2) &\qquad& 
h_0^{'} = -H^2(\mu)/(2\pi^2 T^2),\nonumber\\
h_1=H^2(\mu)/(2\pi^2) &\qquad&
h_1^{'}=-H^0(\mu)/(4\pi^2 T^2),\nonumber\\
h=3h_0[h_0+\mu^2 h_1] &\qquad&
h^{'}=3h_0^{'}[h_0+\mu^2 h_1]+ 3h_0[h_0^{'}+h_1/T^2+\mu^2 h_1^{'}]
\eearr
with $\mu=M_\pi/T$.
Also
$b_{eff}=b-\frac{0.6T}{\pi^3 F^4 M_{\pi}}$,
$b_{eff}^{'}=-\frac{1}{\pi^3 F^4 M_{\pi}^2}
\left(\frac{5}{16}-\frac{0.3 T}{M_\pi}\right)$ and
$a^{'}=\frac{2a}{m_{\pi}^2}
+\frac{3}{32\pi F^2}(1-\frac{35 m_{\pi}^2}{32 \pi^2 F^2})$. 
The constant $c\simeq 0.9$ and
$F_{\pi}/F=1.057\pm 0.012$ with $F_{\pi}=93 MeV$\cite{chpt}.
The constants $a$ and $b$ are related to the S-wave and D-wave
$\pi$-$\pi$ scattering lengths respectively\cite{chpt}.
Finally the functions $H^n(\mu)$ are given as\cite{roh}
\be
H^n(\mu)=\int_0^\infty \frac{x^n dx}{\sqrt{x^2+\mu^2}}
\frac{1}{e^{\sqrt{x^2+\mu^2}}-1}.
\ee

We have extracted the temperature dependence of the condensate as in 
Eq.(\ref{qqchp}) for low temperatures. For temperatures close to $T_c$,
the critical behaviour is that of O(4) spin model in three
dimension\cite{gopal} and has also been seen in lattice QCD
simulations\cite{karsch}. The order parameter here is given as
$\frac{\langle \bar q q\rangle_T}{\langle \bar q q\rangle_{T=0}}
= (1-\frac{T}{T_c})^\beta$, where $\beta=0.39$\cite{lman}. 
We have taken $T_c=150$~MeV\cite{lman}. 
The two regions are joined smoothly and the result is
shown in Fig.~1(a). This result is fitted with
Eq. (\ref{qqrat}) to determine $S(T)=R(0)/R(T)$, which is plotted in
Fig. ~1(b). We shall use it to calculate the quark
propagator and the hadronic correlation functions.

\section{Quark Propagation in Thermal Vacuum}
In the calculation of correlators, quark propagators enter in
a direct manner and hence it is instructive to study aspects
of the interacting propagator in some detail\cite{propcor}. 

The equal time interacting quark Feynman propagator
in the condensate vacuum  is given as $
S_{\alpha \beta}(\vec x)   =  \left< \frac{1}{2} \left[
 \psi_{\alpha}^{i}(\vec x),
             \bar \psi_{\beta}^{i}(0) \right] \right>$,
which at finite temperature reduces to
\begin{eqnarray}
S(\vec x, T)
&=&\frac{1}{2}\frac{\delta^{ij}}{(2\pi)^3}\int e^{i \vec k . \vec x}
\cos\!2\theta[\sin\! 2 h - \vec \gamma \cdot \hat k \cos\! 2h]d\vec k.
\label{prop}\\
&=& \frac{i}{4 \pi^2} \frac{\vec\gamma \cdot \vec x}{x^2}
[I_1(x)-I_2(x)]
  +  \frac{1}{4 \pi^2} \frac{I_3(x)}{x} 
\label{prop2}
\end{eqnarray}
where, 
\bearr
I_1(x) &=& \int_{0}^{\infty } k\; \left( \cos k x - \frac{\sin k
x}{kx} \right) \cos\!2\theta dk,\\
I_3(x)&=&\int_{0}^{\infty } k\; \sin k x \cos\!2\theta 
e^{-R^2(T) k^2} dk,\\
I_2(x) &=& \int_{0}^{\infty } k\; \left( \cos k x - \frac{\sin k
x}{kx} \right) \cos\!2\theta
   \frac{ e^{-R^2(T) k^2}}{1+(1-e^{-R^2(T) k^2})^{1/2}} dk,
\eearr
\noindent
with $x=|\vec x|$, $k=|\vec k|$. 

The free massive propagator, which can be derived from $S(\vec x,T)$ 
by the substitutions $\sin 2f(k)=m_q/\epsilon$ and 
$\cos 2f(k)=k/\epsilon$, is given as
\begin{equation}
S_0(m_q,\vec x,T)  = \frac{1}{(2 \pi)^2} \frac{1}{x} \left[
m_q\left( m_q  K_1(m_q x) -2 I_5(x) \right)
- \; i \; \frac{\vec \gamma \cdot \vec x}{x}
\left( m_q^2 K_2(m_q x)+2I_6(u) \right) \right]  
\label{masfree}
\end{equation}
\[
{\rm where} \;I_5(x)= \int_{0}^{\infty } 
\frac{k}{\epsilon} \sin (kx) \sin^2 \theta dk,
\qquad
I_6(x)=\int_{0}^{\infty}\frac{k^2}{\epsilon}\; 
\left( \cos k x - \frac{\sin k x}{kx} \right) \sin^2\theta dk.
\]
$K_1(m_q x)$ and $K_2(m_q x)$ are the first and second order 
modified Bessel functions of the second kind respectively.
 
In Fig.~2
 we plot the two components Tr $ S(\vec x,T)$ and Tr $(\gamma
\cdot \hat x) S(\vec x,T) $ of the propagator for massless
interacting quarks given by Eq.\ (\ref{prop2}) at 
$T=$ 0 MeV, $T=$ 100 MeV  and $T=$ 135 MeV,
corresponding to the chirality flip and non-flip components 
considered by Shuryak and Verbaarschot \cite{shuqprop}.
The normalisation is discussed in our earlier work\cite{propcor}

To compare with the constituent quark models with an effective
constituent mass, we have also plotted the behaviour of free
massive quark propagator with masses 100 MeV, 200 MeV and
300MeV. In the chirality flip part, the propagator in the
condensate medium starts from zero, consistent with zero quark
mass at small distances, attains a maximum value of about 250
MeV at a distance of about 0.9 fm and then falls off gradually.
Further the interacting propagator overshoots the massive
propagators after about 0.6 fm. We also see that with increasing 
temperature, the chirality flip component has a lower peak and the
position of the peak shifts towards higher distances indicating the
decrease of the dynamical mass with temperature. 

In the chirality non flip part, the interacting propagator
starts from 1, again consistent with the behaviour expected
from asymptotic freedom. But at larger separation it falls
rather fast indicative of an effective mass of the order of
150 MeV. These features are qualitatively similar to that of 
the quark propagator at zero temperature\cite{propcor,shuqprop},
though quantitatively there are differences.
Also, the non-flip component 
falls faster with increase of temperature.

Clearly therefore, similar to the situation at zero temperature, whereas
a constituent quark description is adequate to describe the behaviour
of the chirality nonflip part of the propagator, it is not so for the 
chirality flip part.

\section{Meson Correlation Functions}
In our earlier work, we noted that phenomenology 
of correlation functions necessitated introduction of irreducible four
point structure (or fluctuations of the condensate fields) in vacuum
\cite{plb}.
In fact, the meson correlation functions were different from 
square of the two point function (propagator) and the
difference could be expressed in terms of the four point function. 
The expression of the meson correlation function at zero temperature
defined in our earlier work\cite{plb} can be extended to finite
temperatures as, 
\be
R(\vec x,T)=Tr \left[ S(\vec x,T) \Gamma^{'} S(-\vec x,T) \Gamma \right]
+ Tr \langle |\left[ \Sigma(\vec x) \Gamma^{'} \Sigma(-\vec x)
\Gamma \right]|\rangle_T
\label{totalcor}
\ee
Where 
$ J(x) = \bar \psi_{\alpha}^{i}(x) \Gamma_{\alpha \beta }  
\psi_{\beta}^{j}(x)$, is a generic meson current with
$\Gamma$ being a $4 \times 4$ matrix $(1, \gamma_{5}, 
\gamma_{\mu}$  or $\gamma_{\mu} \gamma_{5} )$,  
$ x \; $is a four vector;
$\alpha$ and $\beta$ are spinor indices;
$i$ and $j$ are flavour indices. The field $\Sigma (\vec x)$
is the condensate fluctuation field introduced in Ref.\cite{plb}
to include four point irreducible structures in QCD vacuum..

Thus, at finite temperature the correlator (Eq.~\ref{totalcor}) 
is now the square of the interacting equal time
{\em thermal} propagator plus the four point contribution at finite
temperature. 
The thermal quark propagator was obtained in the earlier section. 
We keep the structure of the fields $\Sigma (x,T)$ the same as
for zero temperature\cite{plb}. 

\bearr
\Sigma_{\alpha \beta} (\vec x) &=& \Sigma_{\alpha \beta}^V(\vec x) +
\Sigma_{\alpha
  \beta}^S(\vec x) \\
  &=& \mu_1^2 \; (\gamma^i \gamma^j)_{\alpha \beta}\;
\epsilon_{ijk}\;
  \phi^k(\vec x) + \mu_2^2 \; \delta_{\alpha \beta}  \; \phi(\vec x)
\label{sigma}
\eearr
where the first term corresponds to vector fluctuations and the
second to
scalar.  $\mu_1$ and $\mu_2$ in the above equations are dimensional
parameters which give the strength of
the fluctuations and $\phi^k(\vec x)$ and $\phi(\vec x)$ are vector
and scalar fields such that,
 with $|\Omega\rangle$ as the ground state of QCD, we have
\be
\langle \Omega | \phi^i(\vec x) \phi^j(0) | \Omega \rangle = \delta^{ij} g_V(\vec
x) ; \quad
\langle \Omega | \phi(\vec x) \phi(0) | \Omega \rangle = g_S(\vec x)
\ee
At finite temperature, the functions $g_V$ and $g_S$ will be
temperature dependent. We do not know how to calculate it except
for a general property that the effect of the four point structure
should decrease with temperature. We take here a simple ansatz for 
the temperature dependence of $g_V$ and $g_S$,
\be
g_{S,V}(x,T)=\left(\frac{\langle \bar q q\rangle_T}{\langle \bar q
q\rangle_{T=0}}\right)^2 \; \; g_{S,V}(x,T=0)
\label{gsvt}
\ee

Similar to  calculations at zero temperature, we shall consider the
ratio of the physical
correlation function to that of massless noninteracting quarks
at zero temperature  given as
$$ R_{0}(x) = Tr\left[ S_{0}(x)
\Gamma^{'}S_{0}(-x)\Gamma\right] $$.
The normalised correlation functions thus defined as
\begin{equation}
C(\vec x , T)=\frac{R(\vec x, T)}{R_0(\vec x)}
\label{cxt}
\end{equation}
are plotted in Figure ~3 for the pseudoscalar and vector
channels.

As expected (on physical grounds) the amplitude of the correlator
decreases with increasing temperature. The peak of the vector
correlator shifts towards the right after $T=0.9T_c$. We might
remind ourselves that the position of the peak of the correlator is
inversely proportional to the mass of the particle in the relevant
channel\cite{neglecor}.

The spatial hadronic correlators have been used to extract the 
hadronic screening masses and widths at finite
temperature\cite{shuhotcor}. 
To extract the hadronic properties at finite temperature, we use a
phenomenological parameterisation as is usually done in sum rule
calculations\cite{hatsuda,hatsuda2}. We may note here however,
that the phenomenological inputs are not available at finite
temperature. The correlators are parameterised with the mass, decay
width and the coupling of the particle to the vacuum, all three
parameters being temperature dependent.
We first express the correlator in terms of spectral density function.
\be
R_{ph}(\vec x) = \int_0^{\infty} ds \; \frac{\sqrt{s}}{4\pi^2 x} K_1
(\sqrt{s}x) \; \rho (s)
\label{corspec}
\ee
Then we use the following phenomenological parameterisation for 
the spectral density function\cite{hatsuda,hatsuda2},
\be
\rho^V(s) = 3 \lambda_\rho^2 \delta (s-M_{\rho}^2) + 
\frac{3s}{4 \pi^2}\tanh \left[\frac{\sqrt{s}}{4T}\right]\theta (s-s_o)  
+ T^2 {S_\rho}\delta (s)
\label{specvec}
\ee
\be
\rho^P(s) = \lambda_\pi^2 \delta (s-M_\pi^2) + \frac{3s}{8
\pi^2} \tanh \left[\frac{\sqrt{s}}{4T}\right]
\theta (s-s_o) 
\ee
where $\lambda$ is the coupling of the bound state to the current,
$M$ is mass of the bound state and $s_0$ is the threshold
for continuum contributions. 
The last term in  Eq.~(\ref{specvec})
is the scattering term for soft thermal
dissociations (mainly through pions), which exists only at finite
temperature\cite{hatsuda}. This term is given as
\be
S_\rho=\lim _{|\vec p|\rightarrow 0}
\frac{1}{2\pi}\int_0^{|\vec p|^2}
d\omega^2\int_v^\infty x^2 \left(n(\frac{|\vec p|x-\omega}{2T})
-n(\frac{|\vec p|x+\omega}{2T})\right)
\ee
The derivation of the above expression is slightly tricky and
we have given it in the appendix.
Following Ref.~\cite{hatsuda} we take $S_\rho\approx \frac{T^2}{9}$.
       
The mass, threshold and coupling are then extracted 
such that the correlators as obtained from Eq.~\ref{corspec} agree
with the normalised correlation functions as calculated by us
(Fig.~\ref{corall})\cite{plb}. This is done for each temperature.
The results are plotted in Fig.~\ref{pscapar} for the pseudoscalar
channel and in Fig.~\ref{vecpar} for the vector channel. The
results are also shown in Table~\ref{tab1}.

\section{ Summary and Conclusions}
As can be seen from Fig.~3, with increase in
temperature, the correlation functions have a lower peak indicating
lack of correlations  with temperature. In the vector channel
the mass of the $\rho$ meson appears to decrease beyond 120 MeV. The 
threshold for the continuum also decreases around the same temperature.
The behaviour with temperature of these quantities is qualitatively similar
to that found by Hatsuda {\it et al}\cite{hatsuda2}.
We have also plotted the temperature dependence of the coupling of the
boundstate to the current which decreases with temperature but rather slowly
as compared to mass or the threshold for the
continuum. The temperature dependence of these parameters can be
used to calculate the lepton pair production rate from $\rho$ in the
context of ultra relativistic heavy ion collision experiments to
estimate vector meson mass shift in the medium. 

In the pseudoscalar channel the mass remains almost constant till the
critical temperature whereas the thershold and the coupling decrase with
the temparature \cite {hatptp}. We may note here that in the 
pseudoscalar channel, the contribution to the correlation function
 mostly comes from the fluctuating fields and the temperature behaviour 
as taken in Eq.(\ref{gsvt})  essentially does not shift the position of 
the peak whereas the magnitude of the correlator decreases. That is 
reflected in the above behaviour of the parameters in 
the pseudoscalar channel. We may note here that similar behaviour of
pion mass becoming almost insensitive to temperature below the critical 
temperature was also observed in Ref.\cite{hatptp} where correlation functions
were calculated in a QCD motivated effective theory namely the 
Nambu- Jona Lasinio model.

We would like to add here that the present analysis will be valid 
for temperatures below the critical temperature. Above the critical
temperature there have been calculations essentilly using finite
temperature perturbative QCD in random phase approximations \cite{parikh}. 
However, in the region above $T_C$,
nonperurbative features have been seen to exist from studies
in lattice QCD simulations\cite{lman}. 
In view of this, one possibly has to do a 
hard thermal loop calculation where a partial resummation is 
done\cite{pisarski}.
\section{acknowledgement}
The present work was initiated when one of the authors (HM) was visiting
Department of Physics, University of Bielefeld. He would like to thank
the Physics Department there for providing the facilities 
and Alexander von Humboldt Foundation, Germany for a fellowship during 
that period.

\appendix
\section{}
Here we shall derive the scattering term $S_\rho$. This may be
calculated by considering the imaginary part of the longitudinal
correlator for space like four momenta and can be written as
\be
\rho_l^s (\omega, \vec p)
= \frac{Im\;\Pi_{00}}{|\vec p^2|}
\ee
which is explicitly written as\cite{abrik}
\bearr
\rho_l^s (\omega , \vec p) &=& 2 \times \frac{(2\pi)^4}{|\vec p^2|}
\; \int 
\frac{d^3 k_1}{2E_1 (2\pi)^3} \frac{d^3 k_2}{2E_2 (2\pi)^3}
|\langle \pi (\vec k_1)|J_0|\pi (\vec k_2) \rangle |^2\\
 &\times&
\delta (\omega -E_1 + E_2) \delta^{(3)} (\vec p - k_1 + k_2) 
(n_2-n_1).
\label{specdef}
\eearr
Here $E_1=\sqrt{\vec k_1^2+m_\pi^2} \; ; \; E_2=\sqrt{\vec k_2^2+m_\pi^2}$
and $n_i \equiv n(E_i)$ is the Bose distribution function for pions.

In general the expectation of a vector current with respect to a
pion state is given as\cite{don}
\be
\langle \pi ( k_1)|J_\mu|\pi ( k_2) \rangle
= ( k_1 +  k_2)_\mu G_\pi (p)
\ee
where, $p=k_1-k_2$ and  $G_\pi (p)$ is the pion form factor  with 
$G_\pi (0) =1 $
Substituting this in Eq.~(\ref{specdef}) and integrating over $k_2$
we obtain
\be
S_\rho (\omega , \vec p)= 2 \times \frac{2}{(2\pi)^2}{4 |\vec p^2|}
\; \int 
\frac{d^3 k_1}{E_1 }(G_\pi(p))^2\delta(\omega-E_1+E-2)(2E_1-\omega)^2(n_2-n_1)
\end{equation}
with $\vec k_2=\vec p-\vec k_1$. Next, since the delta function above
contribute to space like ($p^2<0$) region we  write it as
$$\delta(\omega-E_1+E_2)=2 E_2\delta((\omega-E_1)^2-E_2^2)\theta(-p^2)$$
To simplify further, we may change the integration over three momentum
 $\vec k_1$ to the integration over energy $E_1$ and the angle
 $\cos\theta_{\vec p, \vec k}$. Performing the integration over 
angles restricts
the lower limit of the enegry integral $E_1$ as $E_{1min}=
{\displaystyle{\frac{1}{2}}}(\omega+|\vec p|v)$, where,
 $v=(1-\frac{4m_\pi^2}{p^2})^{1/2}$. Thus we have
\be
\rho_l^s(\omega,\vec p)=\frac{1}{4\pi} |\vec p|^3\int_{E_{min}}^\infty
G_\pi^2(p)(2E_1-\omega)^2(n_2-n_1)
\ee
with $E_2-E_1-\omega$. Next, defining the varible $x$ through
$E_1=\frac{1}{2}(\omega+|\vec p|x)^{1/2}$, leads to
\be
\rho_l^s(\omega,\vec p)=\frac{1}{8\pi} 
\int_v ^\infty dx x^2 n(\frac{|\vec p|x-\omega}{2T})
G_\pi^2(p)(2E_1-\omega)^2(n_2-n_1)
\ee
We shall consider the longitudinal form factor $S_\rho(\omega,\vec p)$ 
in a frame which is at rest with respect to the medium 
which implies that $\vec p \rightarrow 0$.
In this limit the constraint $0 < \omega < \vec
p^2$ also forces $\omega$ to approach zero. However the above 
integral becomes increasingly large as $\vec p \rightarrow 0$
such that the integrated quantity of $S_\rho (\omega, \vec p)$
within the phase space for $\omega$  remains finite. Thus we first
integrate over this region with $\vec p$ finite and then take the
limit $\vec p \rightarrow 0$. Thus let
\be
I= \lim_{|\vec p| \to 0}
\int_0^{|\vec p|^2} d\omega^2 \rho_l^s (\omega, \vec p)
=\frac{S_\rho}{2\pi}
\label{scdef}
\ee
so that $\rho_l^s(\omega, \vec p)$ effectively becomes a delta
function. Thus the spectral density reduces to
\be
\lim_{|\vec p| \to 0} \rho_l^s (\omega, \vec p)
=\delta (\omega^2) \frac{S_\rho}{2\pi}
\ee

 We also note that
there arises no ambiguity from the pion form factor as
 $G_\pi (p=0)=1$.
Now the integral $I$ can be written as 
\be
I= \frac{1}{8 \pi}
\lim_{|\vec p| \to 0}
\int_0^{|\vec p|^2} d\omega^2 
\int_{v}^\infty dx \; x^2 
\left[n((|\vec p|x-\omega)/2T)-n((|\vec p|x+\omega)/2T)\right]
\ee
We change the integration variables\cite{samir} by putting 
$\omega=|\vec p |\lambda$ and 
$x=\sqrt{1+\frac{y^2}{|\vec p |^2(1-\lambda^2)}}$. 
Hence the spectral density function can be written as
\be
I = \frac{1}{4\pi}
\lim_{|\vec p| \to 0}
\int_0^1  d\lambda \;\lambda 
\int_{2m_\pi}^\infty \frac{x y \; dy}{(1-\lambda^2)^2} 
\left[n(\frac{|\vec p|x-\omega}{2T})-n(\frac{|\vec p|x+\omega}{2T})\right ]
\label{I2}
\ee
In the limit of $|\vec p|\to 0$, we may Taylor expand the difference of
the distribution functions in the square bracket of eq.(\ref{I2})
and have
$$
\left[n(\frac{|\vec p|x-\omega}{2T})-n(\frac{|\vec p|x+\omega}{2T})\right ]
\approx -\frac{2 x|\vec p|^2(1-\lambda^2)^2}{y^2}\frac{dn}{d\lambda}.$$
Substituting back in eq. (\ref{I2})and performing an integration by parts for
$d\lambda$ integration we have
\be
I=\frac{1}{2\pi}\int_0^1d\lambda \int_{2m_\pi}^\infty dy
n(\frac{y}{2T\sqrt{1-\lambda^2}})y.
\ee
In the limit of vanishing pion mas we have $I=2\pi T^2/9$ so that
$S_\rho=T^2/9$.

\begin{figure}
\epsfbox[78 271 524 648]{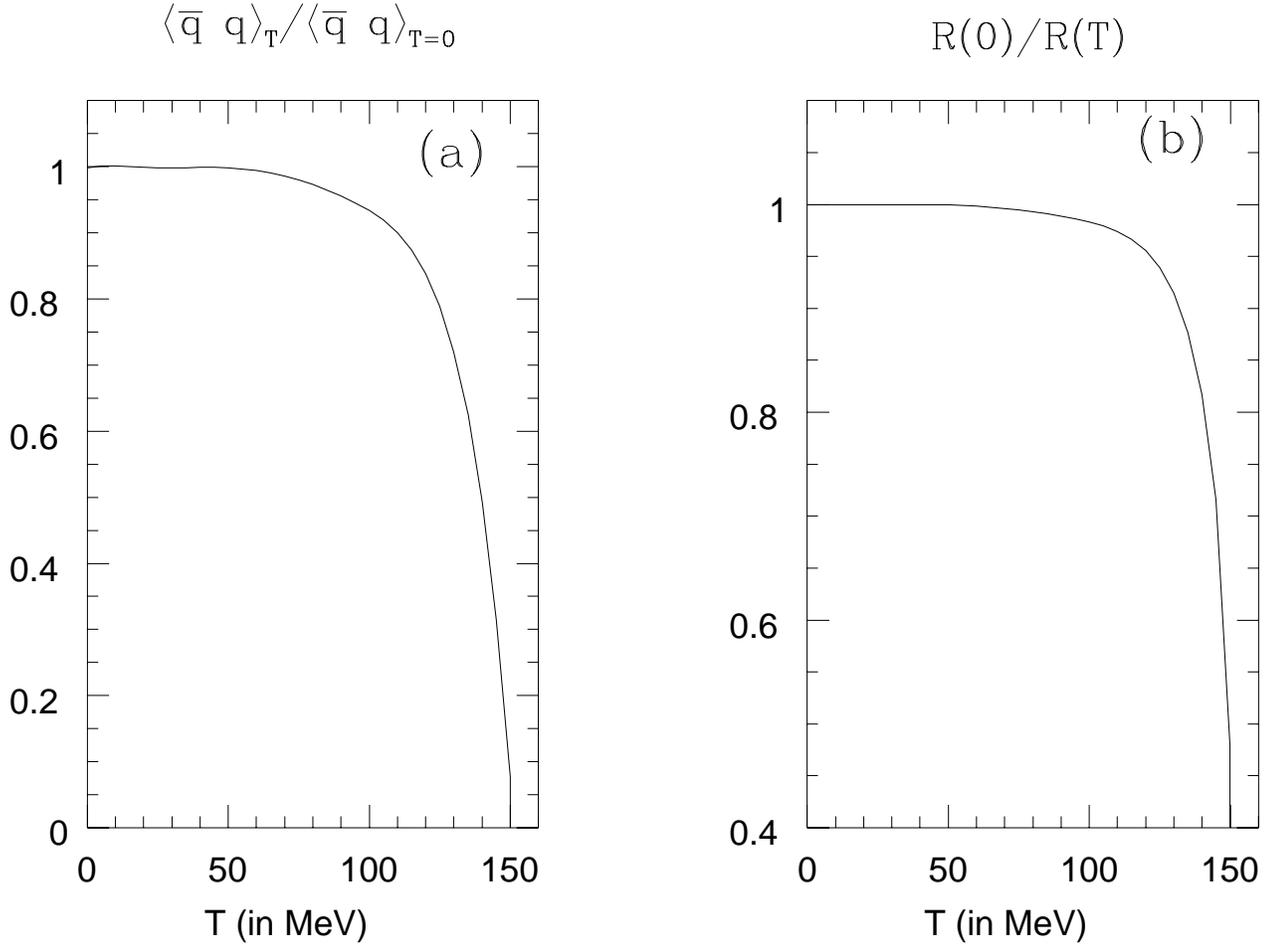} 
\caption{Figure (a) shows quark condensate at finite 
temperature normalised to that at zero temperature obtained from CHPT 
and Lattice. Figure (b) shows R(0)/R(T) as determined from Fig (a).}
\end{figure}

\begin{figure}
\hspace*{0.8in}
\epsfysize=10cm
\epsfbox[78 271 524 648]{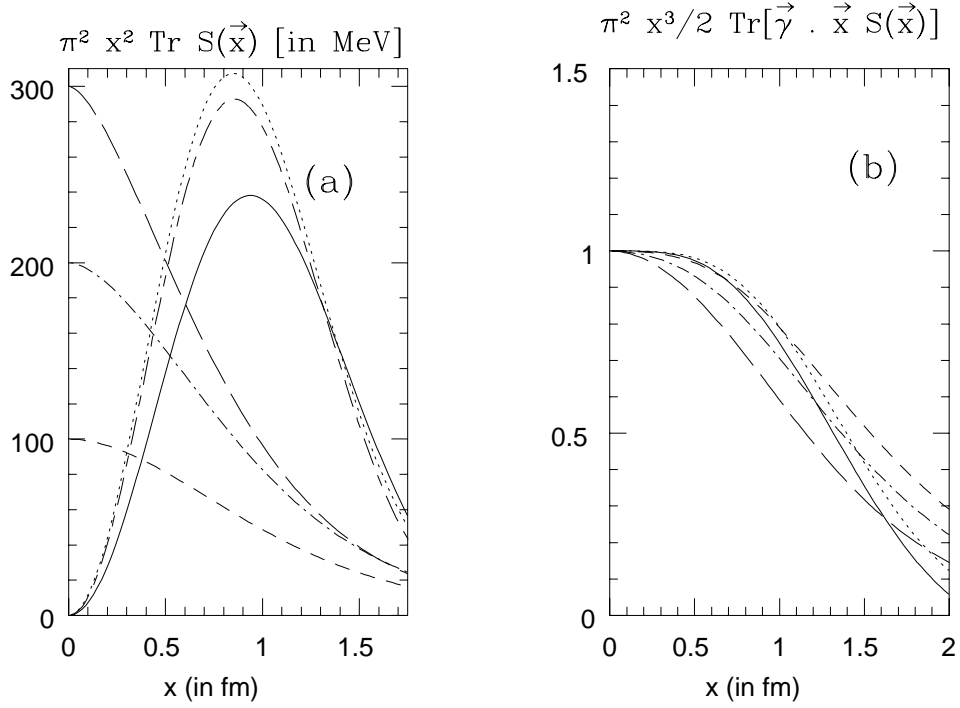}
\caption{The two components of the thermal quark propagator, 
(a) Tr(S) and (b) Tr[($\vec \gamma \cdot \hat x$)S] versus the distance
x(in fm). 
The three lines, dot, short dash-long dash and solid corresponds to
massless quark interacting propagator S(x,T) at temperatures of 0,
100 and 135 MeV respectively. The three lines,
short dashed, dot-short dashed and long dashed correspond to a
massive free propagator with a mass of 100, 200 and 300 MeV,
respectively at T=135 MeV.}
\end{figure}

\begin{figure}
\hspace*{-0.8in}
\epsfbox[22 360 531 653]{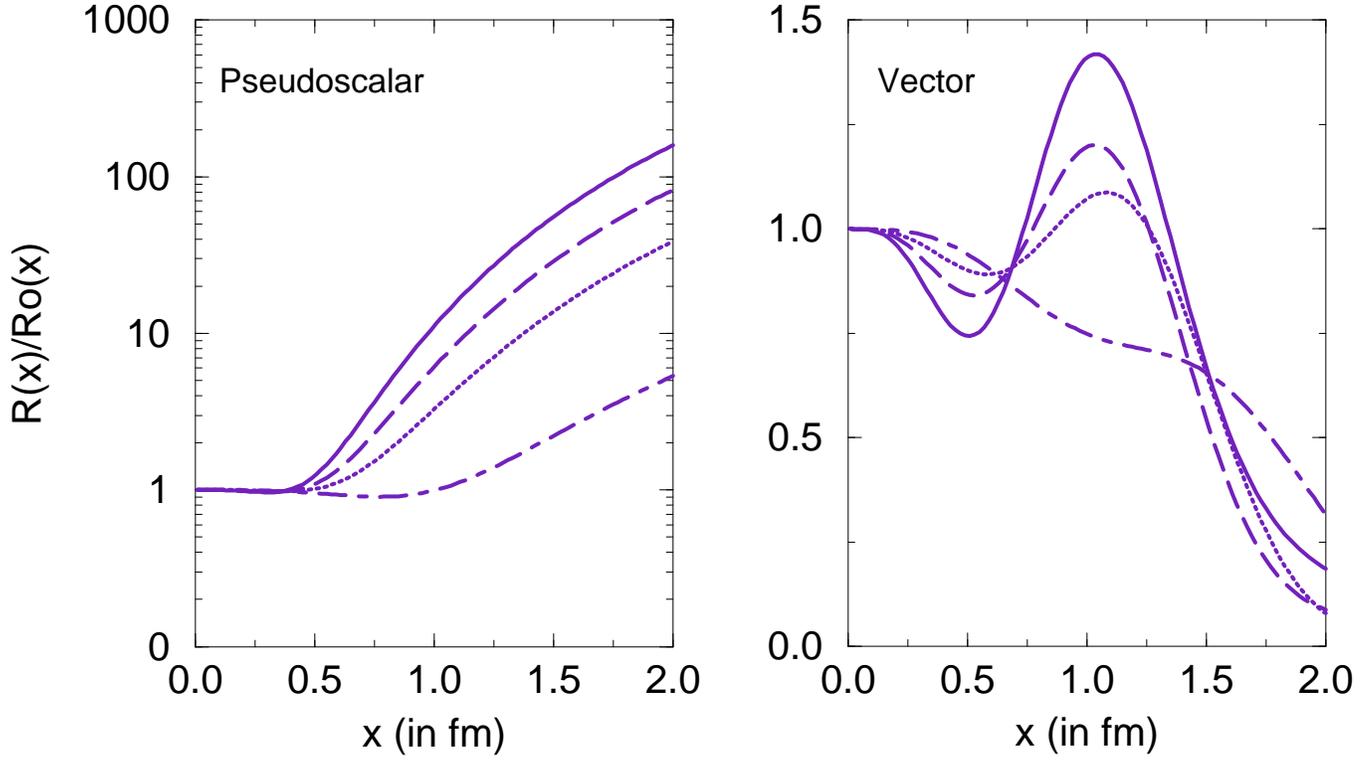}
\caption{The ratio of the meson correlation functions 
at finite
temperature to the correlation functions for noninteracting massless
quarks at zero temperature
$\displaystyle \frac{R(x,T)}{R_{0}(x,T=0)}  $, vs. distance x (in fm). 
The solid, dashed, dotted and dot-dashed lines correspond to temperatures 
$T=$0 MeV, $T=$130 MeV, $T=$140 MeV and $T=$148 MeV respectively.}
\label{corall}
\end{figure}

\begin{figure}
\epsfysize=20cm
\epsfbox[53 30 525 700]{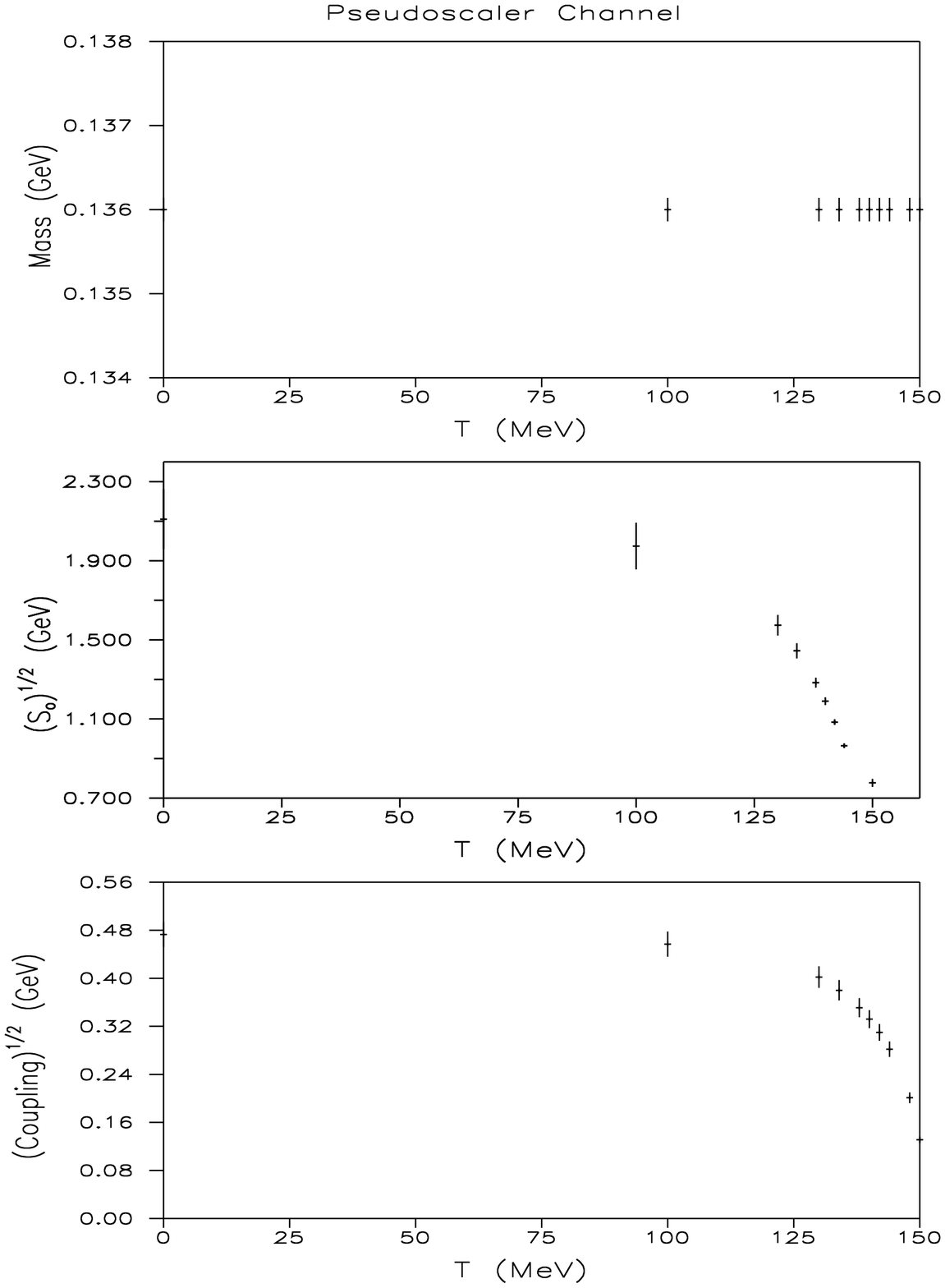}
\caption{The temperature dependence of mass, threshold (S$_0$) and
coupling for the pseudoscalar channel. T$_c=$150 MeV.
The vertical lines represent the errors obtained while fitting}
\label{pscapar}
\end{figure}

\begin{figure}
\epsfysize=20cm
\epsfbox[45 30 525 700]{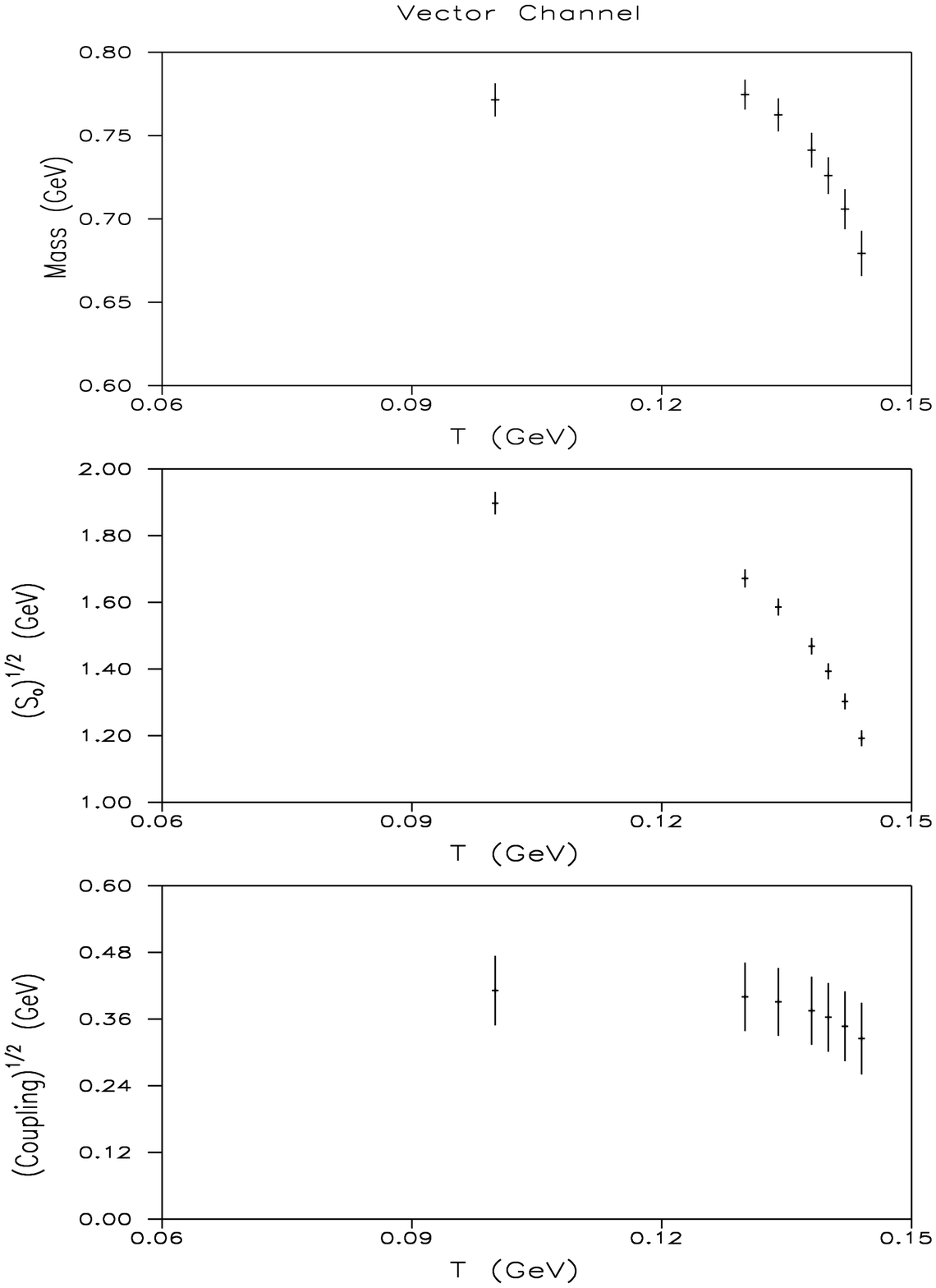}
\caption{The temperature dependence of mass, threshold (S$_0$) and
coupling for the vector channel. T$_c=$150 MeV.
The vertical lines represent the errors obtained while fitting}
\label{vecpar}
\end{figure}

\begin{table}
\caption{Fitted parameters}
\vspace*{0.3in}
\begin{tabular}{lllll}
CHANNEL & Temp.(MeV) & M (GeV) & $\lambda$ & $\sqrt{s_0}$(GeV) \\
\tableline
%****************************************************
Vector & 0 & 0.780 $\pm$ 0.005 & (0.420 $\pm$ 0.041 GeV)$^2$
& 2.070 $\pm$ 0.035 \\
 & 100 & 0.771 $\pm$ 0.001 & (0.411 $\pm$ 0.062 GeV)$^2$
& 1.897 $\pm$ 0.033 \\
 & 130 & 0.774 $\pm$ 0.001 & (0.402 $\pm$ 0.061 GeV)$^2$
& 1.672 $\pm$ 0.027 \\
 & 134 & 0.762 $\pm$ 0.001 & (0.391 $\pm$ 0.061 GeV)$^2$
& 1.586 $\pm$ 0.026 \\
& 138 & 0.741 $\pm$ 0.001 & (0.375 $\pm$ 0.061 GeV)$^2$
& 1.468 $\pm$ 0.024   \\
& 140 & 0.726 $\pm$ 0.001 &  (0.363 $\pm$ 0.062 GeV)$^2$ 
& 1.393 $\pm$ 0.024 \\
& 142 & 0.706 $\pm$ 0.001 &  (0.347 $\pm$ 0.063 GeV)$^2$ 
& 1.303 $\pm$ 0.024 \\
& 144 & 0.679 $\pm$ 0.001 &   (0.325 $\pm$ 0.064 GeV)$^2$
& 1.192 $\pm$ 0.024 \\
\hline  
%****************************************************
Pseudoscalar & 0 & 0.136 $\pm$ 0.00014 & (0.473 $\pm$ 0.021 GeV)$^2$
& 2.110 $\pm$ 0.152 \\
 & 100 & 0.136 $\pm$ 0.00014& (0.457 $\pm$ 0.021 GeV)$^2$
& 1.974 $\pm$ 0.118 \\
 & 130 & 0.136 $\pm$ 0.00014 & (0.402 $\pm$ 0.018 GeV)$^2$
& 1.574 $\pm$ 0.052 \\
 & 134 & 0.136 $\pm$ 0.00014 & (0.380 $\pm$ 0.017 GeV)$^2$
& 1.445 $\pm$ 0.038 \\
 & 138 & 0.136 $\pm$ 0.00014 & (0.351 $\pm$ 0.016 GeV)$^2$
& 1.284 $\pm$ 0.025 \\
 & 140 & 0.136 $\pm$ 0.00014 & (0.332 $\pm$ 0.015 GeV)$^2$
& 1.190 $\pm$ 0.020 \\
 & 142 & 0.136 $\pm$ 0.00014 & (0.310 $\pm$ 0.014 GeV)$^2$
& 1.084 $\pm$ 0.014 \\
 & 144 & 0.136 $\pm$ 0.00014 & (0.282 $\pm$ 0.013 GeV)$^2$
& 0.965 $\pm$ 0.011 \\
 & 148 & 0.136 $\pm$ 0.00014 & (0.201 $\pm$ 0.009 GeV)$^2$
& 0.970 $\pm$ 0.020 \\
 & 150 & 0.136 $\pm$ 0.00014 & (0.131 $\pm$ 0.006 GeV)$^2$
& 0.777 $\pm$ 0.020 \\
\hline  
\end{tabular}
\label{tab1}
\end{table}

\end{document}